# Melting of gold by ultrashort laser pulses: Advanced two-temperature modeling and comparison with surface damage experiments

Sergey A. Lizunov[a,b], Alexander V. Bulgakov[a,b], Eleanor E. B. Campbell[c,d], Nadezhda M. Bulgakova[a,*]

[a] HiLASE Centre, Institute of Physics of the Czech Academy of Sciences, Za Radnicí 828, 25241 Dolní Břežany, Czech Republic

[b] S.S. Kutateladze Institute of Thermophysics SB RAS, 1 Lavrentyev Ave., 630090 Novosibirsk, Russia

[c] EaStCHEM, School of Chemistry, University of Edinburgh, David Brewster Road, Edinburgh EH9 3FJ, United Kingdom

[d] Department of Physics, Ewha Womans University, Seoul 03760, Republic of Korea

* Corresponding author: bulgakova@fzu.cz

**Abstract**

The ultrafast laser-induced solid-liquid phase transition in metals is still not clearly understood and its accurate quantitative description remains a challenge. Here we systematically investigated, both experimentally and theoretically, the melting of gold by single femto- and picosecond near-infrared laser pulses. Two laser systems with wavelengths of 800 and 1030 nm and pulse durations ranging from 124 fs to 7 ps were used and the damage and ablation thresholds were determined for each irradiation condition. The theoretical analysis was based on two-temperature modeling. Different expressions for the electron-lattice coupling rate and contribution of ballistic electrons were examined. In addition, the number of free electrons involved in the optical response is suggested to be dependent on the laser intensity and the influence of the fraction of involved electrons on the damage threshold was investigated. Only one combination of modelling parameters was able to describe consistently all the measured damage thresholds. Physical arguments are presented to explain the modeling results.

# 1. Introduction

Ultrashort-pulse laser action on different types of materials has been utilized as a highly valuable scientific research tool for several decades and presently it gains growing recognition for industrial applications due to tremendous improvement in precision of material processing that reduces the time and cost of manufacturing [1,2]. For achieving desired results in material processing, the damage threshold (DT) is one of the key parameters, which is usually defined as the onset of laser-induced material melting [3-6]. It is a well-defined parameter, which can be measured with a high precision, thus serving as a reference laser fluence for material processing applications and giving valuable information for understanding the material heating dynamics under the laser action. For the latter, the DT measurements provide a well-determined experimental reference that is crucial for developing adequate models that can predict material behavior under different laser exposure conditions.

The interaction of ultrashort laser pulses with metals is typically described by the two-temperature model (TTM) [4,7-9]. This consists of two heat flow equations; one for the conduction electrons, that absorb the laser radiation, and one for the lattice. The two subsystems are connected via an energy relaxation term. The results of the modeling for the spatiotemporal evolution of the electron and lattice temperatures and the maximum heating of the lattice strongly depend on the thermophysical parameters of the free electron gas and the electron-lattice coupling parameter, whose temperature-dependent values are still poorly established. Most simulations reported in the literature are based on the TTM under the assumption of a linear dependence of the heat capacity and the thermal conductivity of electrons on their temperature while the electron-lattice coupling factor is considered to be constant. These assumptions are, however, only applicable to situations where the electron temperature $T_e$ is much lower that the Fermi temperature, $T_F$ [4,8,10]. A deep insight into fundamental aspects of the electron gas behavior under non-equilibrium laser heating conditions can be obtained based on the Boltzmann equation [11,12], which however is difficult to use for practical application in view of the large computer resources required. There are many attempts to improve the TTM description using more sophisticated dependences of the free electron gas properties [8,9,13-16], in particular, utilizing parameters calculated based on the density of states theory developed in [17] under the assumption of equilibrium conditions. However, the actual parameters of the free electron gas in metals are expected to lie in between the two extremes, resulting in a more complicated behavior under the highly non-equilibrium conditions obtained under ultrafast laser action [18]. Taking into account the additional complexity of a theoretical description connected with a dynamic change of the optical properties [9,15], a fraction of free electrons involved in the optical response of metals [15,19], and a possible role of ballistic electrons in heat transfer

[4,20], the TTM allows several degrees of freedom to adjust modeling results to experimental data.

A large body of the experimental data on laser-induced melting and ablation of different metals has been reported, often showing contradicting results for the DT for the same material, in particular for gold, an extensively studied metal where the reported DT values vary almost by an order of magnitude under similar irradiation conditions [4,21-26]. This can be explained by at least two factors, different methods used for determination of the melting threshold and the surface properties of irradiated samples, which can be very different in terms of surface quality and presence of impurities. Indeed, the damage (melting) threshold determination is based either on microscopic inspection of "fingerprints" left on the surface by local melting [21,23-26] or by *in situ* detection of sample optical properties like reflectivity or scattering [4,22,27]. The first method appears to be more precise assuming that a fairly high resolution is used for imaging. The reflectivity/scattering measurement method requires an accurate localization of a probe beam or imaging point within the irradiation spot and a sufficiently thick molten depth of the order of the skin layer thickness or higher. The surface roughness can significantly change the surface reflectivity due to increased surface scattering [9,28,29].

To gain insight into the influence of different factors listed above on the damage threshold of metals, we have performed a comprehensive experimental and theoretical study of the DT for the case of gold, which can be extended to other materials, under three different irradiation conditions. Experiments were performed with the *same gold samples* at a laser wavelength of 800 nm with pulse duration of 124 fs and at a wavelength of 1030 nm with pulse durations of 260 fs and 7 ps. The aim was to identify a unified modelling approach which would describe all experiments with the same material parameters. Numerical simulations were performed based on the two-temperature model (TTM) with different sets of the material parameters. We have investigated the possible role of ballistic electrons, the fraction of free electrons involved in the optical response and the reported values of the electron lattice coupling and have identified a set of TTM parameters, which provides good agreement with all three experimental conditions.

## 2. Experiment and measured thresholds

Gold samples were irradiated under three different conditions using two ultrashort laser systems with Gaussian pulses, a Ti:sapphire laser (Coherent, wavelength $\lambda$ = 800 nm, pulse duration (FWHM) $\tau$ = 124 fs) and a Yb:KGW-based laser (PHAROS, Light Conversion, $\lambda$ = 1030 nm, $\tau$ = 260 fs and 7 ps). Two types of polycrystalline gold samples (both of 99.999% purity) were used in the experiment, bulk gold for 800-nm pules and a commercial thick film on silicon wafers (300 nm Au film thickness, 10 nm Cr adhesion layer) for 1030-nm pulses. All the samples had a high surface quality. The sample characterization by atomic force microscopy (AFM, NanoMan-V) resulted in the root mean square (RMS) surface roughness $\sigma = 10.2$ nm for the bulk gold and $\sigma = 4.7$ nm for the film. The reflectivities $R$ of the sample surfaces were measured by ellipsometry and by direct detection of reflected unfocused laser beams and found to be, for the bulk sample, $R = 0.95$ at 800 nm and $R = 0.978$ at 1030 nm for the film. The measured reflectivity of the film corresponds well with literature data [30] for ideally polished gold surfaces ($R = 0.973$ and $0.98$ at 800 and 1030 nm, respectively). For the bulk samples, the absorption of the laser pulse was approximately twice the value of what would be expected for an ideal surface. It should be noted that, for rough surfaces with a reflectivity close to mirror quality ($\sigma \ll \lambda$), the following expression can be used to evaluate the surface roughness from specular reflectance $R_{rough}$ [31,32]

$$R_{rough} = R \, \exp\left(-\left(\frac{4\pi\sigma}{\lambda}\right)^2\right), \tag{1}$$

where $R$ is the reflection coefficient for the ideally polished surface. Applying the expression (1) for our conditions gives a surface roughness of ~9 nm that is in good agreement with the AFM measurements. In the modelling, the optical properties adjusted to the measured data were used as described below.

The laser beams were focused on the targets at normal incidence by N-BK7 glass lenses (focal distance 25 cm for $\lambda = 800$ nm and 16 cm for $\lambda = 1030$ nm) into circular spots of radius $\omega_0$ (1/$e^2$ criterion) of ca. 58 µm for $\lambda = 800$ nm and 28 µm for $\lambda = 1030$ nm (exact spot sizes were measured in every experiment). The pulse energy $E_0$ was adjusted using a $\lambda/2$ plate in combination with a Glan polarizer to give a peak laser fluence $F_0 = 2E_0/\pi\omega_0^2$ in the range 0.5–5 J/cm$^2$. Most of the experiments were performed in ambient air, although some measurements were repeated in vacuum in order to be certain that the air environment does not affect the threshold determination. The studied fluence range is well below the threshold for laser air ionization (air breakdown) [31]. However, the presence of the highly reflective gold samples can significantly, up to fourfold, increase the local laser intensity in front of the target surface due to interference of the reflected beam with the incident pulse tail [32]. This can potentially result in air ionization in the near-surface region with absorption and scattering of the laser beam by the air plasma and a corresponding decrease in the laser energy reaching the surface. Therefore, for the shortest studied pulse duration of 124 fs, the experiments were also carried out in a vacuum chamber (base pressure 10$^{-6}$ mbar) under the same irradiation conditions as in air and the transmission losses through the chamber window optical elements were taken into account in the laser fluence calculations. All experiments were performed under single-pulse conditions.

The laser-produced spots on the gold surface were examined using an optical microscope in Nomarski mode (Olympus BX43). Typical images of spots produced by picosecond pulses at different laser fluences are shown in Fig. 1. At fairly high fluences, the spots consist of two clearly distinguished regions (Fig. 1,*b*,*c*) with an ablation (inner) region surrounded by an outer region with a modified (damaged) surface. The latter is presumably due to surface melting and, for our very smooth surfaces, has a sharp boundary allowing precise determination of the damaged area. At low fluences, below the ablation threshold, only the damaged area is observed (Fig. 1,*a*). Note that, due to the high stability of the used lasers (shot-to-shot deviation in energy less than 0.5%) and high quality of the target surfaces, the single-shot damage and ablation areas are well reproduced from pulse to pulse with an accuracy better than 1%.

The threshold fluence values were determined based on the measured damage and ablation areas as a function of the pulse energy using the standard procedure for Gaussian beams when the damaged/ablated area *S* is related to the pulse energy $E_0$ by [5,33,34]

$$S = \frac{\pi \omega_0^2}{2} \ln(E_0/E_{th}) \qquad (2)$$

where $E_{th}$ is the threshold energy. By plotting the measured *S* value as a function of ln$E_0$ we can obtain both the effective spot radius $\omega_0$ and the threshold fluence $F_{th} = 2E_{th}/\pi\omega_0^2$. Figure 2 shows such plots for spots produced by 800-nm, 124-fs pulses in air (Fig. 2,*a*) and vacuum (Fig. 2,*b*). The data yield good straight line fits in semilog plots confirming the Gaussian profile of the laser beam. The results obtained in air and vacuum are nearly identical and provide the same threshold fluences for both damage and ablation. This indicates that the presence of air has little effect on the gold damaged and ablated areas in the considered fluence range and the obtained threshold values can be considered to be accurate (note that the ablation rates in air and vacuum can be different even near ablation thresholds due to recondensation of ablated material on the surface in air [35] and possible air ionization at the beam center [32]). Figures 3, *a* and *b* show experimental results for femtosecond and picosecond 1030-nm pulses, respectively. The damage and ablation thresholds deduced from these experiments are summarized in Table 1. The indicated uncertainty ranges correspond to twice the standard deviation obtained based on 7-10 independent measurement series (depending on the sample) from the error estimates of the $\omega_0$ and $E_{th}$ values determined from the semilog fits with Eq. (2). The obtained *damage* thresholds were further used as reference values for modelling the ultrafast-laser heating and melting of gold.

**Table 1.** Measured thresholds for gold damage ($F_{th,\ damage}$) and ablation ($F_{th,\ ablation}$) by ultrashort laser pulses.

| λ, nm | τ, ps | $F_{th,\ damage}$, J/cm² | $F_{th,\ ablation}$, J/cm² |
|---|---|---|---|
| 800 | 0.124 | 0.54 ± 0.04 | 1.3 ± 0.1 |
| 1030 | 0.26 | 1.05 ± 0.03 | 1.65 ± 0.05 |
| 1030 | 7.0 | 1.15 ± 0.03 | 1.7 ± 0.05 |

## 3. Details of the modelling approach

The TTM is based on the assumption that an ultrashort laser pulse brings material to a highly non-equilibrium condition so that the temperature $T_e$ of the conduction electrons absorbing the laser radiation becomes much higher than the lattice temperature $T_l$ [36]. The interaction between the electron and lattice subsystems leads to equilibration of their temperature on the picosecond time scale [4,7]. In metals, the heat transfer is usually described in one dimension (1D), toward the material depth. This can be justified by considering that the depth of laser energy penetration achieved by the time equilibration has been reached (tens-hundreds of nanometers) is much smaller than the size of the irradiation spot (typical diameters of irradiated area in experiments with femtosecond lasers are in the range 10-100 µm). Therefore, the 1D equations of the TTM can be written as

$$C_e \frac{\partial T_e}{\partial t} = \frac{\partial}{\partial x}\left(K_e \frac{\partial T_e}{\partial x}\right) - g(T_e - T_l) + S(x,t), \tag{3}$$

$$C_l \frac{\partial T_l}{\partial t} = \frac{\partial}{\partial x}\left(K_l \frac{\partial T_l}{\partial x}\right) + g(T_e - T_l). \tag{4}$$

Here $C_e$, $C_l$, $K_e$, $K_l$ are the heat capacities and the thermal conductivities for electrons and lattice respectively, $g$ is the coupling factor which characterizes the rate of energy exchange between the electrons and the lattice, $t$ is time, $x$ is the coordinate toward the target depth (at the surface $x = 0$), and $S(x,t)$ is the energy source term, which describes the laser energy absorption. Note that the term of the lattice thermal conduction in Eq. (4) (first term on the right) can be neglected due to a small lattice heat transport on sub-nanosecond time scales.

The temporal shape of the laser pulse intensity $I$ on the target surface can be described by the Gaussian function as

$$I(0,t) = \frac{F(1-R)}{\tau}\sqrt{\frac{\ln 2}{\pi}} \exp\left[-4\ln 2 \left(\frac{t-t_m}{\tau}\right)^2\right]. \tag{5}$$

Here $F$ is the fluence of the laser pulse, $R$ the reflection coefficient, which can be strongly dependent on material heating, $\tau$ is the pulse duration (FWHM), and $t_m$ is the time at which the laser intensity reaches its maximum, counted from the beginning of the simulations (here we have chosen $t_m = 3\tau$).

Attenuation of laser radiation toward the metal depth can be described according to the Beer–Lambert law as

$$\frac{dI(x,t)}{dx} = -\alpha(x,t)I(x,t) \qquad (6)$$

where $\alpha(x,t)$ is the absorption coefficient. Note that generally this coefficient can depend on several factors, mainly on the electron temperature, and therefore it can be time- and space-dependent. Thus, we introduce the laser energy absorption in a more general way via Eq. (6), contrary to the widely used formulation where an exponential attenuation of the laser light with a constant $\alpha$ value is used.

The initial temperatures of the electrons and the lattice are set to be 300 K. The boundary condition on the irradiated surface of the metal sample corresponds to the absence of heat flux through the boundary while, deep inside the metal target, the temperature is considered to be constant and equal to the initial value

$$\left(\frac{\partial T}{\partial x}\right)\bigg|_{x=0} = 0, \quad T|_{x=\infty} = 300 \text{ K}. \qquad (7)$$

To describe the heat capacity, the thermal conductivity and the electron-lattice coupling factor of gold, the data obtained by Lin *et al.* based on the density of state theory [17] are used, which are approximated by the following dependences:

$$C_e \text{ [J/(m}^3\text{K)]} = 10^5 \times \left(0{,}7207 - 6{,}7589\,\tilde{T}_e + 50{,}74116\,\tilde{T}_e^2 - 32{,}73731\,\tilde{T}_e^3 + 6{,}44355\,\tilde{T}_e^4\right) \quad (8)$$

$$g \text{ [W/(m}^3\text{K)]} = \begin{cases} 2.6 \times 10^{16}; \; T_e < 2600 \text{ K} \\ 10^{17} \times (0.3905 - 2.02752\,\tilde{T}_e + 6.98567\,\tilde{T}_e^2 - 5.765999\,\tilde{T}_e^3 + \\ \qquad 1.92621\,\tilde{T}_e^4 - 0.23943\,\tilde{T}_e^5); \; T_e > 2600 \text{ K} \end{cases} \quad (9)$$

$$K_e = v_F^2 C_e / 3\nu_e \text{ [W/(m}^3\text{K)]}. \qquad (10)$$

Here $\tilde{T}_e = T_e/10^4$, $v_F$ is the Fermi velocity, $\nu_e = AT_e^2 + BT_l$ is the electron collision frequency where the first and second terms correspond to electron-electron and electron-ion collisions respectively. The *B* coefficient was calculated via Eq. (10) at low temperatures when normally $AT_e^2 \ll BT_l$ and one may consider $C_e = C_0 T_e$ [4] assuming thermal equilibrium with $T_e = T_l$. When the coefficient *B* is determined, the *A* value is calculated by linking Eqs. (13) and (21) of Ref. [37]. Note that special attention is given to match the $\nu_e$ values with optical properties of the studied gold samples at normal conditions. The *B* coefficient for cold samples was verified via the Drude model with some fitting of its value and the plasma frequency $\omega_p$ to the measured reflectivity values at the studied wavelengths. The corresponding parameters are given in Table 2.

**Table 2**. Parameters for gold used in simulations.

| Parameter | Value |
|---|---|
| $v_F$ $[m/s]$ | $1.39 \times 10^6$ |
| $H_m$ $[J/m^3]$ | $1.3 \times 10^9$ |
| $C_l$ $[J/(m^3 K)]$ | $2.5 \times 10^6$ |
| $A$ $[K^{-2} s^{-1}]$ | $1.2 \times 10^7$ |
| $B$ $(800\ nm)$ $[K^{-1} s^{-1}]$ | $5.29 \times 10^{11}$ |
| $\omega_p$ $(800\ nm)$ $[s^{-1}]$ | $1.2 \times 10^{16}$ |
| $B$ $(1030\ nm)$ $[K^{-1} s^{-1}]$ | $4.5 \times 10^{11}$ |
| $\omega_p$ $(1030\ nm)$ $[s^{-1}]$ | $1.29 \times 10^{16}$ |

We note that the coupling factor $g$ given by Eq. (9) is obtained under the assumption of thermal equilibrium of the electrons. However, under ultrafast laser heating, the electron subsystem can deviate from the equilibrium. Therefore, here we perform comparative simulations using the equilibrium $g$ value and the value obtained in the work of Mo *et al*. [18] by visualizing ultrafast laser melting of gold using electron diffraction. Their measurements showed that a constant coupling factor $g = 2.2 \times 10^{16} [W/(m^3 K)]$ provides a good fit to the data obtained near the onset of the gold melting process.

The melting process is simulated as a classical Stefan problem. Upon reaching the melting point in a numerical cell, the temperature is fixed until the additional absorbed energy in this cell becomes equal to the heat of fusion $H_m$. After this, the material in this cell is allowed to be heated above the melting point. The melting threshold is defined as the minimum fluence at which melting is achieved at the surface.

The Drude model is used to describe the optical properties of gold from determining the dynamic behaviour of the reflection and absorption coefficients as

$$\varepsilon = \varepsilon_1 + \iota \varepsilon_2 = \varepsilon_0 - \frac{\omega_p^2}{\omega(\omega + \iota \nu_e)}, \tag{11}$$

$$n = \sqrt{0.5(\varepsilon_1 + \sqrt{\varepsilon_1^2 + \varepsilon_1^2})}, \tag{12}$$

$$k = \sqrt{0.5(-\varepsilon_1 + \sqrt{\varepsilon_1^2 + \varepsilon_1^2})}, \tag{13}$$

$$R = \frac{(n-1)^2 + k^2}{(n+1)^2 + k^2}, \tag{14}$$

$$\alpha = \frac{4\pi k}{\lambda}. \tag{15}$$

Here $\varepsilon$ is the complex dielectric function which depends on the radiation wavelength $\omega$; $c$ is the speed of light; $n$ and $k$ are the refractive index and the extinction coefficient in the media. For normal conditions according to Ref. [30], the $n$ and $k$ values are respectively 0.178 and 5.02 for the wavelength of 800 nm that corresponds to $R = 0.973$ and 0.272 and 7.07 for 1030 nm that corresponds to $R = 0.979$. As mentioned above, for the $T_e$-dependent optical model, the $\omega_p$ and $\nu_e$ values were adjusted to the experimental values of the reflectivity under the normal conditions, $R = 0.95$ at 800 nm for the bulk and $R = 0.978$ at 1030 nm for the film.

Another fundamental question concerning laser energy absorption in metals is the role of so-called ballistic electrons. It is generally assumed that, in gold exposed to ultrashort powerful laser radiation pulses, some of the electrons within the radiation absorbing layer at the surface become highly energetic and transfer ballistically the absorbed energy deeper into the sample, well beyond the nominal absorption depth $l_{ab} = 1/\alpha$ [38]. This leads to a decrease of the local density of the absorbed energy at the irradiated surface and, as a result, the damage threshold may increase. In a number of models, the effect of the ballistic electrons is taken into account in the form of the so-called ballistic length $l_{ball}$, which is assumed to be 100 nm for gold [4] with the value being metal-specific [39]. To account for ballistic electrons, the gold absorption coefficient $\alpha$ is modified to the form $\alpha_{ball} = 1/(l_{ab} + l_{ball})$ [4]. We note that, in such formulations, the overall optical absorption depth is increased dramatically, irrespective of the laser intensity. However, one can expect that the fraction of ballistic electrons and the energy transferred by them toward the sample depth are dependent on irradiation conditions and change with the laser energy flux. In this paper, we have tested both optical models, with and without accounting for the ballistic electrons to verify their effect at laser fluences near the damage threshold.

The final important issue addressed in our modelling concerns the fraction of free electrons involved in the optical response of metals. In several publications, it is stated that only approximately a third of the electrons contribute to the absorption and reflection of the incoming light [15,19]. Here we assume that the number of electrons involved in the optical response to laser irradiation (in other words, the free electron current induced by the field of the electromagnetic wave) can depend on the field intensity. At relatively low laser intensities, this fraction can be rather small while increasing at higher intensities. In the simulations, we systematically varied the fraction of the electrons $\eta$ contributing to the Drude model (Eqs. (11)-(15)) in the range of $\eta = 0.1-1$ via correspondingly scaling the coefficient $A$ in the electron collision frequency $\nu_e$ as $\eta A$. Note that all conduction electrons were considered to contribute to the heat conductivity ($\eta = 1$ in Eq. (9)).

Four variants of the model were applied in attempts to describe the laser-induced damage threshold of gold under the three different irradiation conditions of the experiments (see Section

2) as summarized in Table 3, namely with and without ballistic electrons and with two different values of the electron-lattice coupling factor. For each model variant, the detailed calculations have been performed with different fractions of the free electrons involved in the optical response.

**Table 3.** The model variants used in the simulations (MV)

| Model number | Ballistic electrons | Electron-lattice coupling factor | Model symbol in Fig. 4 |
|---|---|---|---|
| MV1 | Yes | Constant [18] | ■ |
| MV2 | Yes | Polynomial, Eq. (8) | ● |
| MV3 | No | Constant [18] | ◆ |
| MV4 | No | Polynomial, Eq. (8) | ▲ |

**4. Comparison of numerical calculations with the experiment and discussion**

The majority of TTM modelling results known in the literature for metallic samples are reported for cases of radiation exposure at one wavelength and, as a rule, for one pulse duration. The experiments carried out in this work made it possible to systematically simulate the damage thresholds of gold and to compare the modelling results with the experimental data obtained under different irradiation conditions for samples with high-quality surfaces having well-defined optical properties. The simulation results are presented in Fig. 4. They report the damage thresholds obtained numerically using the model variants specified in Table 3. The damage thresholds are given as a function of the fraction of the conduction electrons involved in the optical response to the laser radiation. We remind that in the model, all the conduction electrons are considered to contribute to the thermal conductivity.

It is obvious that MV1 and MV4 do not describe adequately the experimental damage threshold values. The only exception is for the femtosecond pulse at a wavelength of 1030 nm (Fig. 4,*b*), provided that all free electrons are fully involved in the optical response, which contradicts the available studies [15] and appears to be rather unlikely. Regarding MV2, it yields a reasonable description of the damage thresholds for all three experimental regimes only if it is assumed that all or almost all conduction electrons contribute to the optical response that again raises doubts for relatively low laser intensities. Additionally, according to the experiments on visualization of the gold melting dynamics [18], at laser intensities near the melting threshold, the electron-lattice relaxation is slowed down and its rate is significantly lower than the values calculated in the framework of the equilibrium density functional theory [17]. Thus, the physical bases of MV2 raise certain doubts. Therefore, we can conclude that MV3, which includes a value for the electron-lattice coupling factor corresponding to measurements [18] and which assumes a negligible contribution of ballistic electrons, seems to be most appropriate.

According to MV3, in the regimes near the melting threshold, the fraction of free electrons contributing to the optical response of gold to laser radiation is small, of the order of 0.2-0.3 (Fig. 4), which is reasonably consistent with the data available in the literature (see [15] and references therein). From the physical point of view, it can be speculated that, at relatively low intensities of polarized laser light, mostly the electrons with momenta in the direction of polarization are involved in the electric current in the direction of polarization. The higher the laser intensity, the more electrons can be involved in the laser-induced electric current and, hence, in the optical response to the laser field. Furthermore, we can hypothesize that, at high laser intensities, all electrons participate in the oscillating current activated by a strong laser field, provided that the laser pulse is long enough for the electrons to receive momenta along the polarization direction in the collision events.

Concerning the contribution of ballistic electrons, the question of their influence on the melting threshold has not yet been sufficiently studied. The introduction of a ballistic length into the metal absorption coefficient, regardless of the intensity of laser pulses, seems to be physically ungrounded. This assumption implies that all electrons absorbing laser radiation are immediately involved in the ballistic energy transfer, leading to a broadening of the absorption zone in gold by more than an order of magnitude. We note that experimental studies of the ballistic electron transport in metals using femtosecond laser irradiation are usually performed at very low fluences, typically of the order of 1 mJ/cm$^2$ that is much lower than typical melting thresholds (more than two orders of magnitude for gold). Ballistic transport can be observed at distances smaller than the characteristic scattering length of electrons [20] or, in other words, at low electron collision frequencies, here $\nu_e = AT_e^2 + BT_l$. Under the action of ultrashort laser pulses with fluences well below the melting threshold, the electrons, although nonequilibrium, still have low collision frequencies with the condition $AT_e^2 \ll BT_l$ and, hence, a relatively large scattering length at which the ballistic transport can be observed.

For gold, the situation overturns when the electrons are suddenly heated to several thousands of Kelvin and higher where $AT_e^2 > BT_l$ and $\nu_e$ starts to grow quadratically with respect to $T_e$. Near the melting threshold, the maxima of the electron temperature achieved during the laser pulse at the sample surface are ~16 000 K for fs laser pulses at both wavelengths and in excess of 10 000 K for the ps pulses, implying that the ballistic range should be strongly reduced or even completely suppressed by electron-electron collisions. This supports the assumptions of MV3, which provides the best agreement of the simulation results with the measured damaged thresholds for all three experimental regimes. Finally, based on our simulations and the above analysis, we can conclude that, at laser fluences near and above the melting threshold, the energy transport in gold is diffusive rather than ballistic and that the fraction of electrons involved in the optical response should increase with increasing intensity of the laser pulse. However, a detailed understanding of the full significance of these two statements provides important fundamental topics for further research.

## 5. Conclusions

In this paper, a comprehensive model has been developed targeted on advanced comparisons of the modelling results with experimental data. It is based on the two-temperature model and takes into account the dynamic change of the optical response of metallic material. A wide range of calculations for different irradiation parameters was performed for gold applying different modelling assumptions used in the literature (different descriptions of the electron-lattice coupling rate, accounting for the input of ballistic electrons to the energy transport, partial involvement of electrons in the optical response at moderate laser pulse intensities).

We also presented new experimental data for gold damage and ablation thresholds for ultrashort laser pulses. The data were gathered for identical gold samples with well characterized mirror-like surfaces irradiated by femto- and picosecond laser pulses at two wavelengths, 800 and 1030 nm. By comparing the experimental data and the numerical results, we found a variant of the model which describes well the conditions of gold irradiation by ultrashort laser pulses. It has been shown that the contribution of ballistic electrons to absorbed energy transport should be disregarded at relatively high laser fluences typical for laser material processing regimes. These conclusions, however, call for further detailed studies to clarify their generality.

**Acknowledgements**

The authors thank Dr. J. Bonse for providing the bulk gold target. This work was supported by the European Regional Development Fund and the state budget of the Czech Republic (project BIATRI: No. CZ.02.1.01/0.0/0.0/15_003/0000445). S. A. L. and A. V. B. also acknowledge financial support from the Russian Foundation for Basic Research (project No. 19-38-90203).

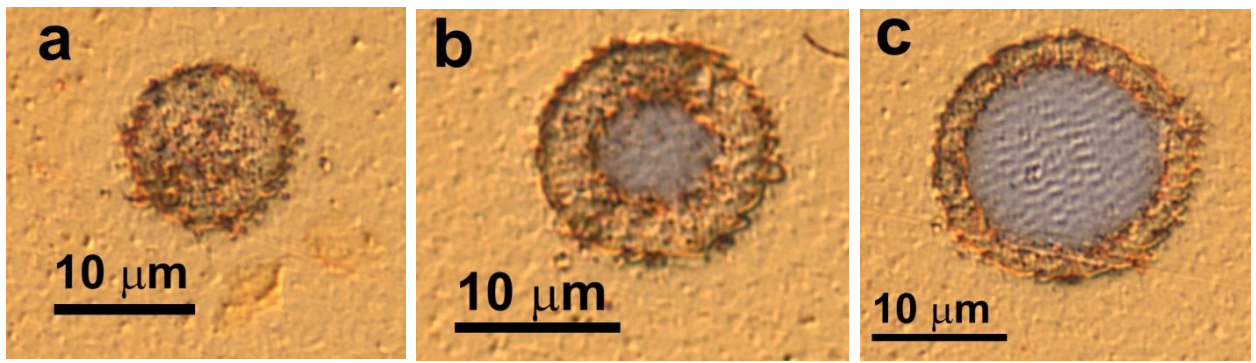

**Fig. 1.** Images of spots produced on the thick film gold surface by single 1030-nm 7-ps laser pulses at fluences of 1.5 (*a*), 2.0 (*b*) and 2.8 J/cm$^2$ (*c*).

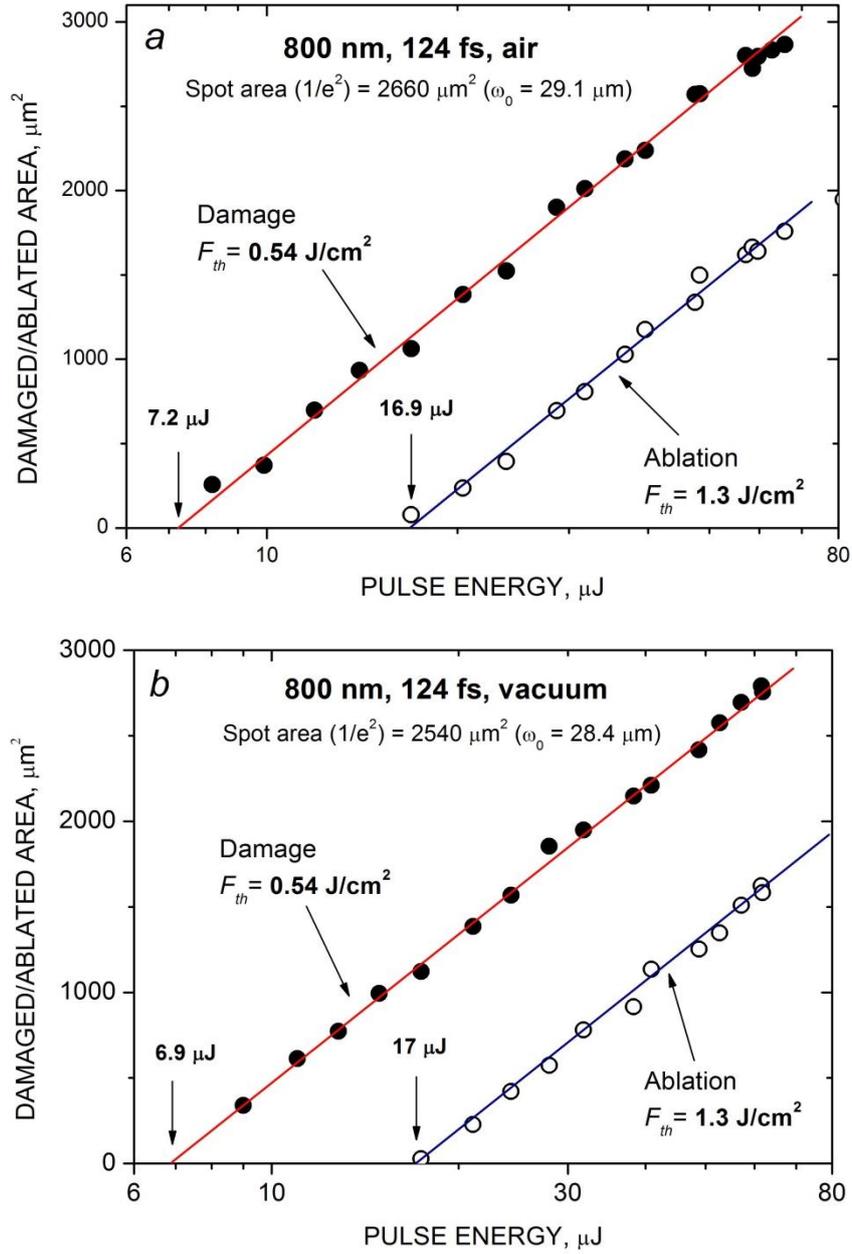

**Fig. 2.** Damaged and ablated areas of spots produced on the gold surfaces by 800-nm 124-fs laser pulses as a function of laser pulse energy in air (*a*) and in vacuum (*b*). The lines represent the least-square fits according to Eq. (1). The measured threshold energies, fluences and effective spot radii are indicated.

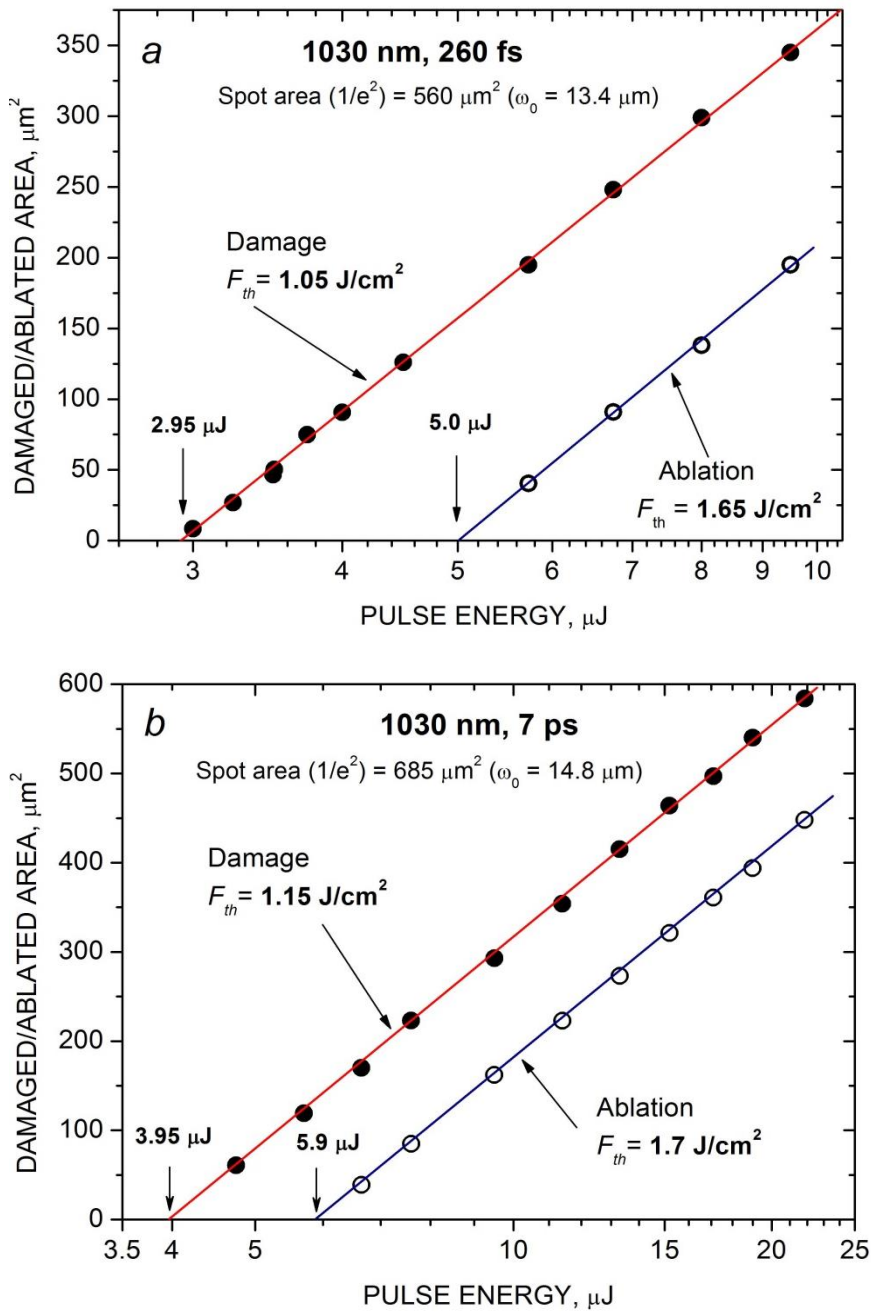

**Fig. 3.** Same as in Fig. 2 for spots produced on thick gold films by 1030-nm laser pulses with duration of 260 fs (*a*) and 7 ps (*b*).

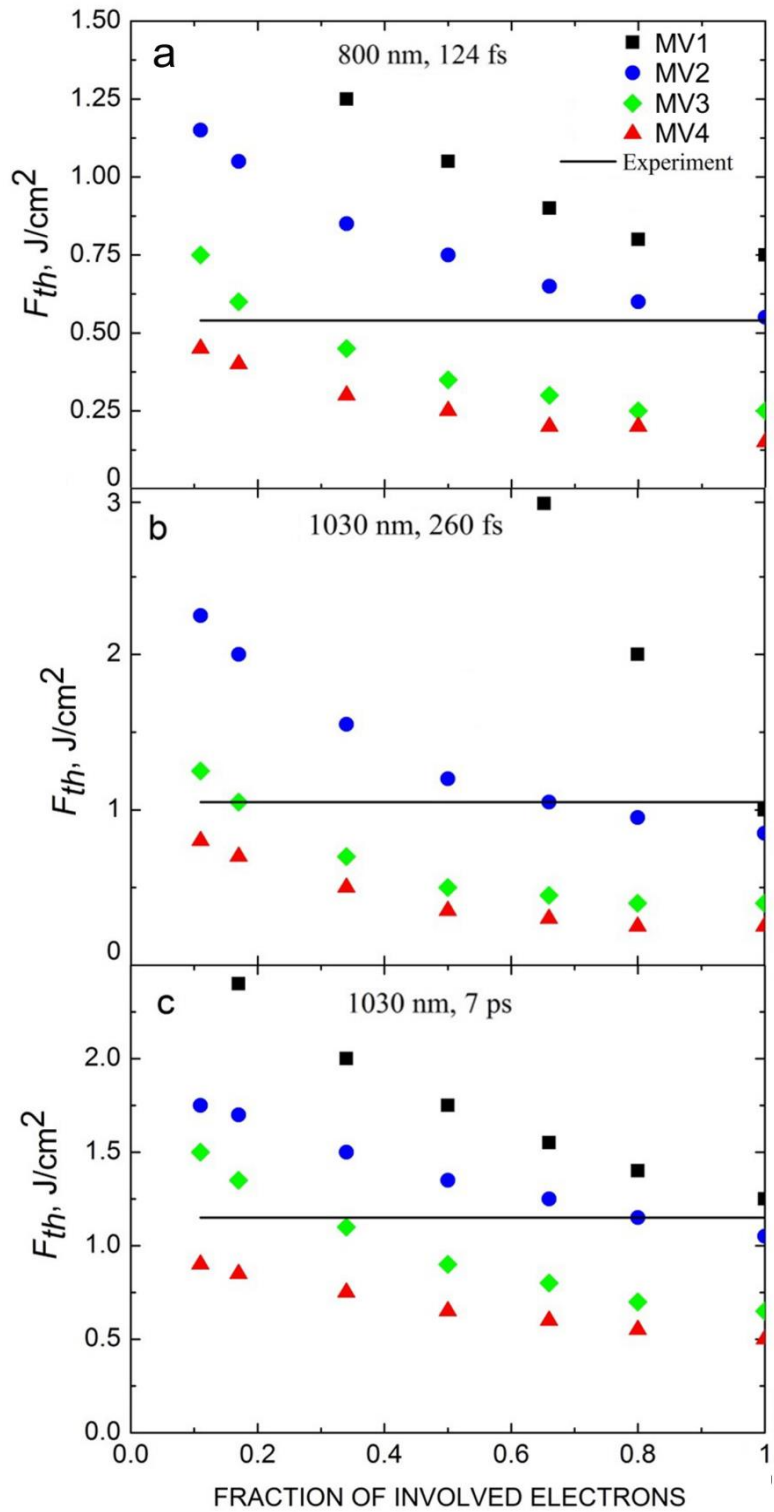

**Fig. 4.** Calculated damage thresholds of gold as a function of the fraction of the conduction electrons involved in the optical response. (a) Femtosecond pulse at 800 nm (b) Femtosecond pulse at 1030 nm. (c) Picosecond pulse of 1030 nm. Solid lines show the measured thresholds. The legend refers to the theoretical models as specified in Table 3.